\documentclass[9pt]{article}
\usepackage{spconf,amsmath,graphicx,hyperref}
\usepackage{placeins}      
\usepackage{subcaption}    

\title{On the Optimality of Rate Balancing for Max-Min Fair Multicasting}
\name{Sadaf Syed, Wolfgang Utschick, Michael Joham}
\address{School of Computation, Information and Technology, Technical University of Munich, Germany}
\newcommand{\bfa}{{\boldsymbol a}}
\newcommand{\bfb}{{\boldsymbol b}}

\newcommand{\bfd}{{\boldsymbol d}}

\newcommand{\bfh}{{\boldsymbol h}}

\newcommand{\bfw}{{\boldsymbol w}}

\newcommand{\bfA}{{\boldsymbol A}} 
\newcommand{\bfB}{{\boldsymbol B}}

\newcommand{\bfD}{{\boldsymbol D}}

\newcommand{\bfH}{{\boldsymbol H}}

\newcommand{\Tm}{\mathrm{T}}

\newcommand{\Hm}{{\mathrm{H}}}

\usepackage{cite}
\usepackage{amsmath,amssymb,amsfonts}
\usepackage{graphicx}
\usepackage{textcomp}
\usepackage{xcolor}
\def\BibTeX{{\rm B\kern-.05em{\sc i\kern-.025em b}\kern-.08em
    T\kern-.1667em\lower.7ex\hbox{E}\kern-.125emX}}

\usepackage{afterpage}
\usepackage{tikz}

\usepackage{lipsum}
\usepackage{fancyhdr}

\usetikzlibrary{%
	arrows,%
	automata,%
	backgrounds,%
	calc,%
	patterns,%
	plotmarks,%
	positioning,%
	shapes.geometric,%
	shapes,%
	snakes,%
	decorations.pathmorphing,%
	decorations.shapes,%
	graphs,%
	chains,%
	arrows.meta, %
}

\usepackage{cite}
\usepackage{amsmath,amssymb,amsfonts}
\usepackage{lipsum}
\usepackage{graphicx}
\usepackage{subcaption}
\usepackage{hyperref}
\usepackage{textcomp}
\usepackage{xcolor}
\def\BibTeX{{\rm B\kern-.05em{\sc i\kern-.025em b}\kern-.08em T\kern-.1667em\lower.7ex\hbox{E}\kern-.125emX}}

\usepackage{tikz}
\usepackage{pgfplots}
\pgfplotsset{compat=newest}
\pgfplotsset{plot coordinates/math parser=false}
\newlength\figH  
\newlength\figW  
\usetikzlibrary{quotes,arrows.meta}
\usepackage{physics}
\usepackage{amssymb}
\usepackage{amsmath}
\usepackage{amsthm}
\usepackage{float}
\usepackage{algorithm}
\usepackage{algpseudocode}
\usepackage{algorithm,caption}

\usepackage[english]{babel}

\usepackage{graphicx}
\usepackage{wrapfig}
\usepackage{caption}
\usepackage{array}
\usepackage{multirow}
\usepackage{arydshln}
\usepackage{color}
\usepackage{xcolor}
\usepackage{tcolorbox}
\usepackage[framemethod=tikz]{mdframed}
\fancypagestyle{IEEEtitlepagestyle}{
    \fancyhf{} 
    \fancyfoot[L]{\vspace{12pt}%
    \footnotesize
    \copyright This work has been submitted to the IEEE for possible publication. Copyright may be transferred without notice, after which this version may no longer be accessible.}
}

\begin{document}
%
\maketitle
\thispagestyle{IEEEtitlepagestyle}

\begin{abstract}
The max–min fair (MMF) multicasting problem is known to be NP-hard. In this work, we analytically derive the optimal solution to this NP-hard problem and establish the equivalence between rate balancing and the optimal MMF multicasting solution under certain conditions. Based on this theoretical insight, we propose a low-complexity algorithm for MMF multicasting that yields closed-form solutions. Simulation results validate our analysis and demonstrate that the proposed algorithm outperforms the state-of-the-art methods while being computationally more efficient. 
\end{abstract}
\section{Introduction}
The ever‐increasing demand for high data rates and spectral efficiency in wireless communication systems has renewed interest in multicasting, where the same data is transmitted to multiple users \cite{multi1, bjornson, bjornson2}. The multicasting protocol is particularly beneficial in applications such as live video streaming, emergency alerts, software updates, and federated learning, where the same message is sent to large groups of users. Multicasting provides an efficient solution by transmitting a single data stream to multiple users over shared time–frequency resources, thereby enhancing the spectral efficiency and mitigating the co-channel interference. A foundational strategy for physical-layer multicasting was presented in \cite{sidiropoulos}, where the max-min fair (MMF) problem was solved to maximise the minimum received signal-to-noise ratio (SNR) among all users. The original MMF multicasting problem is NP-hard, and hence, most state-of-the-art algorithms focus on suboptimal designs that can achieve near-optimal performance \cite{sidiropoulos, bjornson, bjornson2, hunger2007design}. These algorithms are based on semidefinite relaxation (SDR) \cite{sidiropoulos, bjornson}, alternating direction method of multipliers (ADMM) \cite{bjornson2} or heuristic techniques such as \cite{hunger2007design}, which updates the beamforming filters successively while iteratively increasing the SNR. The SDR method is usually implemented using interior point solvers like CVX \cite{grant2014cvx}. Even though the solution obtained by the CVX is sufficiently close to the global optimal solution, the computational complexity of the CVX solvers makes its practical implementation challenging. The MMF multicasting problem is similar to the common rate maximisation problem in rate splitting multiple access (RSMA) \cite{yijiemao, syed2025low, syed2025bilinear}, which also relies on approximations to tackle the NP-hard optimisation problem. The existing algorithms on multicasting or RSMA do not investigate the structure of the optimal solution. Even though \cite{yijiemao} solves the RSMA problem by analysing the Karush-Kuhn-Tucker (KKT) conditions, obtaining the optimal dual variables is not straightforward. In this paper, we employ the fractional programming (FP) approach \cite{fractional1, fractional2} and Lagrangian duality to obtain the optimal solution of MMF multicasting. The proposed algorithm computes the optimal dual variables in closed-form. We analyse the structure of the optimal solution and theoretically prove that under certain conditions, balancing the rates of all users leads to the optimal MMF solution. To the best of our knowledge, this is the first work that establishes the connection between rate balancing and optimal MMF multicasting. The key contributions of this paper are outlined as follows: i) we derive the relationship between rate balancing and the optimal MMF multicasting solution; ii) we propose a low-complexity algorithm with closed-form solutions based on this theoretical insight; and finally, iii) we establish the conditions under which rate balancing is guaranteed to be the optimal solution for multicasting.
\section{System Model and Problem Formulation}
 We consider the downlink~(DL) of a multi-user multiple-input single-output (MISO) system, which is equipped with one base station~(BS) having $M$ antennas and $K$ single-antenna users. The channel realisations are assumed to be perfectly known by the BS. The channel from the BS to the $k$-th user is denoted by $\bfh_{k}\in \mathbb{C}^{M}$ and $n_{k} \sim\mathcal{N_\mathbb{C}}(0, 1) $ denotes the additive white Gaussian noise (AWGN) at the $k$-th user's side. The SNR of the $k$-th user is given by $| \bfh_k^{\Hm} \bfw|^2$ where $\bfw \in \mathbb{C}^{M}$ represents the common multicast precoding vector which should satisfy the DL power constraint $\norm{\bfw}_2^2 \leq P_\text{t}$. The MMF multicasting problem reads as
 \begin{subequations}
\begin{alignat}{2}
&\!\max\limits_{\bfw}     \quad \!\min\limits_{k} | \bfh_k^{\Hm} \bfw|^2    \quad \text{s.t.} \quad      \norm{\bfw}_2^2 \leq P_\text{t}. \tag{P1} \label{P1}
\end{alignat}
\end{subequations}
Note that \eqref{P1} can be equivalently written as
{
\begin{subequations}
\begin{alignat}{2}
&\!\max\limits_{\bfw}     \quad \!\min\limits_{k} \dfrac{| \bfh_k^{\Hm} \bfw|^2}{\bfw ^{\Hm} \bfw}P_t    \quad \text{s.t.} \quad      \norm{\bfw}_2^2 \leq P_\text{t}. \tag{P2} \label{P2}
\end{alignat}
\end{subequations}
}
Now \eqref{P2} can be reformulated using the FP approach \cite{fractional1, fractional2} by introducing auxiliary variables $\beta_k$'s as follows
\begin{subequations}
\begin{alignat}{2}
&\!\max\limits_{\bfw, \forall k:\beta_k}     \: \!\min\limits_{k}   \:  2  \text{Re}\big\{\beta_k^{*}\bfh_k^{\Hm}\bfw\big\} -  \frac{|\beta_k|^2}{P_t} \bfw^{\Hm}\bfw \quad \text{s.t.} \:      \norm{\bfw}_2^2 \leq P_\text{t}. \tag{P3} \label{P3}
\end{alignat}
\end{subequations}
\eqref{P2} and \eqref{P3} are equivalent for the optimal values of $\beta_k$'s 
\begin{align}
  \beta_k^{\star} = \dfrac{\bfh_k^{\Hm}\bfw }{\bfw^{\Hm}\bfw}P_t \:\quad \forall k.
\end{align}
Now \eqref{P3} with optimal $\beta_k^{\star}$ can be reformulated in terms of active and passive users, where $\mathcal{I}$ denotes the set of active users which have the same minimum rate denoted by $z$ and $\mathcal{O}$ denotes the set of inactive users which have higher rates. The problem can be rewritten as
\begin{subequations}
\begin{align}
\max_{\bfw, z} \quad & z \tag{P4}\label{P4} \\
\text{s.t.} \quad 
& z = 2 \Re\!\left\{\beta_k^{*}\bfh_k^{\Hm}\bfw\right\} - \dfrac{|\beta_k|^2}{P_t} \bfw^{\Hm}\bfw, \quad && \forall k \in \mathcal{I}, \label{5a}\\
& z < 2 \Re\!\left\{\beta_k^{*}\bfh_k^{\Hm}\bfw\right\} - \dfrac{|\beta_k|^2}{P_t} \bfw^{\Hm}\bfw, \quad && \forall k \in \mathcal{O} \label{5b}\\
& \|\bfw\|_2^2 \leq P_t. \label{5c}
\end{align}
\end{subequations}
\section{Solution Approach and Equivalence to Rate Balancing}
 In this section, we propose to solve \eqref{P4} based on the KKT optimality conditions. The Lagrangian function for \eqref{P4} is given by
\begin{align}
   &L(\bfw, {\boldsymbol{\lambda}}, z) = z  + \mu \left(P_t - \|\bfw\|_2^2 \right) \nonumber  \\
   &\quad +\sum\limits_{k \in \mathcal{I}} \lambda_k \left(2  \text{Re}\big\{\beta_k^{*}\bfh_k^{\Hm}\bfw\big\} -  \dfrac{|\beta_k|^2}{P_t} \bfw^{\Hm}\bfw - z \right)    \label{22}
\end{align}
where $\mu \geq 0$ is the dual variable corresponding to \eqref{5c} and ${\boldsymbol{\lambda}} = [\lambda_1, \cdots, \lambda_K]^{\Tm} \in \mathbb{R}^K$ is the vector of the dual variables corresponding to \eqref{5a} and \eqref{5b}. Note that $\lambda_k = 0 \: \forall k \in \mathcal{O}$ (inactive constraints) and $\lambda_k \: \forall k \in \mathcal{I}$ can be any real number as it corresponds to the equality constraint in \eqref{5a}. The KKT conditions are next analysed.
\begin{enumerate} 
   \item Taking the derivative of $L$ w.r.t. $z$ results in
   \begin{equation}
    \dfrac{\partial L}{\partial z} = 0 \:\Rightarrow \:  \sum\nolimits_{k=1}^{K}\lambda_k = 1 \label{1}
    \end{equation}
\item The optimal $\bfw^{\star}$ can be computed as
\begin{equation}
\dfrac{\partial L}{\partial \bfw^{*}} = {\bf{0}} \: \Rightarrow \:   \bfw^{\star}  = \dfrac{\sum\nolimits_{k=1}^{K}\lambda_k\beta_k\bfh_k }{\sum\nolimits_{k=1}^{K}\lambda_k |\beta_k|^2/P_t + \mu}  \label{w}
\end{equation}
\item Complementary slackness: 
\begin{align}
   & \lambda_k \left(   2  \text{Re}\big\{\beta_k^{*}\bfh_k^{\Hm}\bfw\big\} -  \dfrac{|\beta_k|^2}{P_t} \bfw^{\Hm}\bfw - z \right) = 0 \quad \forall \:k \label{8}\\
   &\mu \left(P_t - \|\bfw\|_2^2 \right) = 0 \label{9}
\end{align}
\end{enumerate}
Note that at optimality, $\bfw^{\star}$ satisfies \eqref{5c} with equality, and this is ensured by the positive scalar denominator of $\bfw^{\star}$ in \eqref{w}. Hence, $\bfw^{\star}$ from \eqref{w} can also be written as
\begin{align}
    \bfw^{\star} = \zeta\sum\limits_{k=1}^{K}\lambda_k\beta_k\bfh_k \
\end{align}
where $\zeta \geq 0$ ensures $\|\bfw^{\star}\|_2^2 = P_t$. Now considering $\bfH = [\beta_1 \bfh_1, \cdots, \beta_K \bfh_K] \in \mathbb{C}^{M \times K}$, we get
\begin{align}
    \bfw^{\star} = \zeta \bfH {\boldsymbol{\lambda}}. \label{wnew}
\end{align}
Due to \eqref{1} and \eqref{wnew}, the dual function of \eqref{P4} is given by
\begin{align}   L({\boldsymbol{\lambda}}; \bfw^{*}) &= 
   \sum\limits_{k =1}^{K}\lambda_k \left(   2  \text{Re}\big\{\beta_k^{*}\bfh_k^{\Hm}\zeta \bfH {\boldsymbol{\lambda}}\big\} -  |\beta_k|^2\right) \label{24}
\end{align}
since $\lambda_k = 0 \quad \forall k \in \mathcal{O}$.
 In order to compute the optimal $\lambda_k$, we need to minimise the dual function in \eqref{24} w.r.t $\lambda_k$'s, which is given by
\begin{subequations}
\begin{alignat}{2}
&\!\min\limits_{\boldsymbol{\lambda}}    {\boldsymbol{\lambda}}^{\Tm} \bfA{\boldsymbol{\lambda}} - {\boldsymbol{\lambda}}^{\Tm}\bfb   \quad \text{s.t.} \quad  {\boldsymbol{1}}^{\Tm} {\boldsymbol{\lambda}} = 1 \tag{P5} \label{P5}
\end{alignat}
\end{subequations}
where $\bfA = 2 \zeta \: \text{Re} \{\bfB \} \in \mathbf{R}^{K \times K}$, $\bfb = \left[| \beta_1|^2, \cdots, | \beta_K|^2 \right]^{\Tm} \in \mathbf{R}^{K }$, $\bfB = \bfH^{\Hm}\bfH \in \mathbf{R}^{K \times K}$, and ${\boldsymbol{1}}$ is a $K$-dimensional all-ones vector. Since $\bfA$ is a positive semidefinite matrix, \eqref{P5} is convex in ${\boldsymbol{\lambda}}$, its global optimal solution is given by 
\begin{align}
    \boldsymbol{\lambda} = \dfrac{1}{2} \bfA^{-1} \left(\bfb - \eta {\bf{1}} \right)\label{sol2}
\end{align}
if $\bfA$ is invertible (i.e., $\bfA$ is positive definite) and $\eta \in \mathbb{R}$ is the dual variable corresponding to $ {\boldsymbol{1}}^{\Tm} {\boldsymbol{\lambda}} = 1$.  It is to be noted that the formulation of \eqref{P5} inherently takes into account that some $\lambda_k$'s can be zero, and the optimal solution of \eqref{sol2} will take care of that. 
\par {\it{\bf{Theorem 1}}}: If $\bfA$ is invertible, the optimal solution given by \eqref{sol2} delivers balanced rates of all users.
\begin{proof}
For given positive definite matrix $\bfA$ and vector $\bfb$, \eqref{P5} is convex in ${\boldsymbol{\lambda}} $, and its unique optimal solution is given by \eqref{sol2}. We next analyse the solution to \eqref{P4} assuming \eqref{5a} is satisfied for all $K$ users, i.e., the set $\mathcal{O}$ is empty. If all users are active, complementary slackness from \eqref{8} implies 
\begin{align}
  2  \text{Re}\big\{\beta_k^{*}\bfh_k^{\Hm}\bfw\big\} -  \dfrac{|\beta_k|^2}{P_t} \bfw^{\Hm}\bfw - z  = 0 \quad \forall \:k. \label{cs}
\end{align}
Since the optimal $\bfw^{\star}$ will satisfy the power constraint with equality, we have $\norm{\bfw^{\star}}^2_2 = P_t$ at optimality, and accordingly \eqref{cs} can be rewritten as 
\begin{align}
  2  \text{Re}\big\{\beta_k^{*}\bfh_k^{\Hm}\bfw^{\star}\big\} -  |\beta_k|^2 - z  = 0 \quad \forall \:k \label{cs1}
\end{align}
which gives $K$ equations. Plugging in the optimal solution of $\bfw^{\star}$ from \eqref{wnew} in \eqref{cs1}, we get
\begin{align}
    \bfA \boldsymbol{\lambda} - z {\boldsymbol{1}} = \bfb.\label{27} 
\end{align}
Hence, if the optimal dual variable $\boldsymbol{\lambda}$ satisfies \eqref{27}, the resulting solution leads to balanced rates of the users. In order to show the optimality of \eqref{27}, we compare it with \eqref{sol2}, which provides the unique optimal solution of $\boldsymbol{\lambda}$ leading to the optimal solution of \eqref{P5}. Note that for optimal $\bfw^{\star}, $\eqref{24} can also be written as [cf. \eqref{22}]
\begin{align}
   L({\boldsymbol{\lambda}}; \bfw^{*}) &=  {\boldsymbol{\lambda}}^{\Tm} \bfA{\boldsymbol{\lambda}} - {\boldsymbol{\lambda}}^{\Tm}\bfb + z(1 - {\boldsymbol{1}}^{\Tm} {\boldsymbol{\lambda}})\label{28}
\end{align}.
The optimal $\boldsymbol{\lambda}$ from \eqref{28} can be computed as
\begin{align}
\frac{\partial  L({\boldsymbol{\lambda}}; \bfw^{*})}{\partial {\boldsymbol{\lambda}}} = {\bf{0}}\: \Rightarrow \:  \bfA \boldsymbol{\lambda} - z {\boldsymbol{1}} = \bfb \label{29}
\end{align}
which is exactly the same equation as \eqref{27}. This shows that the optimal solution provided by \eqref{sol2} results in the balanced rates of all the users if the matrix $\bfA$ has full rank.
\end{proof}
This means that we can also obtain the same $\boldsymbol{\lambda} $ as in \eqref{sol2} by solving \eqref{cs1}. We get $K$ equations from \eqref{1} and \eqref{cs1} where $K-1$ independent equations arise from \eqref{cs1} by eliminating $z$ as follows
\begin{align}
  2  \text{Re}\big\{\beta_i^{*}\bfh_i^{\Hm}\bfw^{\star}\big\} -  |\beta_i|^2 =  2  \text{Re}\big\{\beta_j^{*}\bfh_j^{\Hm}\bfw^{\star}\big\} -  |\beta_j|^2 \quad \forall \:i \neq j  \nonumber 
\end{align}
Assuming $i=K$ for simplicity, the above equation can be rewritten as
\begin{align}
  2  \text{Re}\left\{\left(\beta_K^{*}\bfh_K^{\Hm} -\beta_j^{*}\bfh_j^{\Hm} \right)\bfw^{\star}\right\}  =  |\beta_K|^2 -  |\beta_j|^2  \: \forall \:j \neq K  \nonumber 
\end{align}
Plugging the expression of the optimal $\bfw^{ \star}$ from \eqref{w} into the above equation, we get
\begin{align}
  2  \text{Re}\{\bfa_j^{\Hm} \sum\limits_{k=1}^{K}\lambda_k \beta_k \bfh_k \} = b_j \left( \mu +  \sum\limits_{k=1}^{K}\lambda_k \dfrac{|\beta_k|^2}{P_t}\right) \: \forall \:j \neq K  \label{eq2} 
\end{align}
where $b_j$ and $\bfa_j$ are given by
\begin{align}
    b_j =  |\beta_K|^2 - |\beta_j|^2 , \quad
     \bfa_j =  \beta_K\bfh_K - \beta_j \bfh_j.  \label{el2}
\end{align}
Hence, we get a linear system of equations in $\lambda_k$'s, given by
\begin{align}
    \bfD \boldsymbol{\lambda} = \bfd \label{lin1}
\end{align}
where the elements of the matrix $\bfD \in \mathbb{R}^{K \times K}$ and the vector $\bfd \in \mathbb{R}^{K}$ are given by $D_{i,j} =  2  \text{Re}\big\{\beta_j \bfa_i^{\Hm} \bfh_j \} - b_i \dfrac{|\beta_j|^2}{P_t} \: \forall \: i = 1:K-1$, $D_{K,j} = 1 \: \forall j$, $d_i = \mu b_i \: \forall \: i = 1:K-1, \: d_K =1.$ $\mu$ in $\bfd$ can be computed by the bisection search which ensures that $\norm{\bfw^\star}_2^2 = P_t$, and if $\bfD$ is invertible, we get
\begin{align}
    \boldsymbol{\lambda} = \bfD^{-1} \bfd \label{sol1}
\end{align}
which gives the same solution as \eqref{sol2}, as proved in Theorem~1. The optimal precoders can be found by inserting \eqref{sol1} into \eqref{wnew}, where $\zeta$ is computed as the scaling factor ensuring that $\norm{\bfw}_2^2 = P_t$.
\par {\it{\bf{Theorem 2}}}: Assuming that the channels of $K$ users are linearly independent vectors, the balanced rates are guaranteed to be the optimal solution of MMF multicasting as long as $K \leq M$. Additionally, for the case when $K > 2 M$, unbalanced user rates will always lead to the optimal solution.
\begin{proof}
Since $\bfB = \bfH^{\Hm}\bfH$ and $\bfH \in \mathbb{C}^{M \times K}$, $\text{rank}\{\bfB\} \leq \text{min} \{ K, M\}$. 
\underline{Case 1} $K \leq M$: If the channels of $K$ users are linearly independent of each other, $\text{rank}\{\bfB\} = K$. Since $\bfA = \zeta \left(\bfB + \bfB^{*} \right)$, $K \leq \text{rank}\{\bfA\} \leq 2 K$. Hence, if $K \leq M$, $\bfA$ is always invertible and we can compute the optimal $\boldsymbol{\lambda}$ by \eqref{sol2} or \eqref{sol1}, leading to the balanced rates of all users as the optimal solution of MMF multicasting. \underline{Case 2} $M < K \leq 2 M$: Here, $\text{rank}\{\bfB\} = M$, and $M \leq \text{rank}\{\bfA\} \leq 2 M$. In this case, rate balancing of all users is the optimal solution only when $\text{rank}\{\bfA\} \geq K$. \underline{Case 3} $K > 2 M$: In this case, $\bfA$ is always rank deficient and rate balancing of all users is not optimal.
\end{proof}
{\it{\bf{Corollary}}}: Since the matrix $\bfD$ is implicitly similar to $\bfA$, the same restrictions on the values of $K$ and $M$ also hold for the invertibility of $\bfD$. Assuming the linear independence of $K$ channels, whenever $\bfD$ is not invertible for the cases $K > M$, we need to find a subset of active users given by $\mathcal{I}$ such that $\bfD_{\mathcal{I}}$ is invertible, giving $ {\boldsymbol{\lambda}}_{\mathcal{I}} =  \bfD^{-1}_{\mathcal{I}} \bfd_{\mathcal{I}}$, and $\lambda_k = 0 \quad \forall k \notin {\mathcal{I}}$. More details on efficiently finding the active set of users will be presented in our future work.
 The proposed method can also be modified to handle the case of collinear channels. Among the set of collinear channel users, we choose the user with the lowest channel norm as the active user, and the rest of the users with higher channel norms are considered inactive. 
\begin{figure*}[!t]
    \centering
    \begin{subfigure}{0.48\textwidth}
        \centering
        \scalebox{0.8}{\definecolor{mycolor1}{rgb}{1.00000,1.00000,0.00000}%
\definecolor{mycolor2}{rgb}{1.00000,0.00000,1.00000}%
\definecolor{mycolor3}{rgb}{0.00000,1.00000,1.00000}%
\begin{tikzpicture}

\begin{axis}[%
width=0.9\figW,
height=\figH,
at={(0\figW,0\figH)},
scale only axis,
xmin=0,
xmax=27,
xlabel style={font=\color{white!15!black}},
xlabel={$\bf{Transmit\:Power\:{\it{P_t}}\:in\:dB}$},
ymin=0,
ymax=80,
ylabel style={font=\color{white!15!black}},
ylabel={$\bf{Min-SNR}$},
axis background/.style={fill=white},
xmajorgrids,
ymajorgrids,
legend style={at={(0,0.8)}, anchor=south west, legend cell align=left, align=left, draw=white!15!black, row sep=-0.05cm, font=\scriptsize},
legend columns=3
]

\addplot [color=blue,  dashdotted,line width=1.0pt, mark=o, mark options={solid, blue}]
  table[row sep=crcr]{%
0	0.150858973947788\\
10	1.50833502708294\\
16.9897000433602	7.5381497011912\\
20	15.0831521508482\\
26.9897000433602	75.3892891956092\\
};
\addlegendentry{CVX Rand \cite{sidiropoulos}}

\addplot [color=red,  line width=1.0pt] 
  table[row sep=crcr]{%
0	0.14758617418806\\
10	1.49073297176453\\
16.9897000433602	7.47005039819902\\
20	15.0567491743233\\
26.9897000433602	75.3309079804671\\
};
\addlegendentry{Algo RB}

\addplot [color=green, dashdotted, line width=1.0pt, mark=o, mark options={solid, green}]
  table[row sep=crcr]{%
0	0.155000662742477\\
10	1.55000665477312\\
16.9897000433602	7.75003328657866\\
20	15.500066564383\\
26.9897000433602	77.500332701537\\
};
\addlegendentry{CVX Relaxed \cite{sidiropoulos}}

\addplot [color=yellow,  line width=1.0pt, mark=o, mark options={solid, yellow}]
  table[row sep=crcr]{%
0	0.0862592888384009\\
10	0.962592888384009\\
16.9897000433602	6.31296444192004\\
20	13.62592888384009\\
26.9897000433602	73.1296444192004\\
};
\addlegendentry{SNR Inc. \cite{hunger2007design}}

\addplot [color=pink, dashdotted, line width=1.0pt, mark=star, mark options={solid, pink}]
  table[row sep=crcr]{%
0	0.130654995840001\\
10	1.3035493495286\\
16.9897000433602	6.52726314188571\\
20	13.065272640501\\
26.9897000433602	73.1256741238833\\
};
\addlegendentry{ADMM \cite{bjornson}}
\end{axis}
\end{tikzpicture}%
}
        \caption{Min-SNR vs $P_t$ for $M = 10, K = 5$}
        \label{fig1}
    \end{subfigure}
    \hfill
    \begin{subfigure}{0.48\textwidth}
        \centering
        \scalebox{0.55}{
%
%
\definecolor{mycolor1}{rgb}{0.00000,0.44700,0.74100}%
\definecolor{mycolor2}{rgb}{0.85000,0.32500,0.09800}%
\definecolor{mycolor3}{rgb}{0.92900,0.69400,0.12500}%
\definecolor{mycolor4}{rgb}{0.49400,0.18400,0.55600}%
\definecolor{mycolor5}{rgb}{0.46600,0.67400,0.18800}%
\begin{tikzpicture}

\begin{axis}[%
width=1.5\figW,
height=1.4\figH,
at={(0\figW,0\figH)},
scale only axis,
bar shift auto,
xmin=0.507692307692308,
xmax=5.49230769230769,
xtick={1, 2, 3, 4, 5},
xlabel style={font=\color{white!15!black}},
xlabel={\textbf{User Index}},
ymin=0.5,
ymax=1.8,
ylabel style={font=\color{white!15!black}},
ylabel={$\bf{SNR}$},
axis background/.style={fill=white},
xmajorgrids,
ymajorgrids,
legend style={at={(0,0.9)}, anchor=south west, legend cell align=left, align=left, draw=white!15!black, row sep=-0.05cm},
legend columns=4
]

\addplot[ybar, bar width=0.123, fill=red, draw=black, area legend] table[row sep=crcr] {%
1	1.60085650365304\\
2	1.60118865961024\\
3	1.60158645699229\\
4	1.60217005334568\\
5	1.60338160536981\\
};
\addplot[forget plot, color=white!15!black] table[row sep=crcr] {%
0.507692307692308	0\\
5.49230769230769	0\\
};
\addlegendentry{Algo RB}

\addplot[ybar, bar width=0.123, fill=green, draw=black, area legend] table[row sep=crcr] {%
1	1.60173197314806\\
2	1.60173197393054\\
3	1.60173197415388\\
4	1.60173197646903\\
5	1.60173197765481\\
};
\addplot[forget plot, color=white!15!black] table[row sep=crcr] {%
0.507692307692308	0\\
5.49230769230769	0\\
};
\addlegendentry{CVX Relaxed \cite{sidiropoulos}}

\addplot[ybar, bar width=0.123, fill=blue, draw=black, area legend] table[row sep=crcr] {%
1	1.60170971772706\\
2	1.60172189719652\\
3	1.60172919873097\\
4	1.60175708249149\\
5	1.60176838293643\\
};
\addplot[forget plot, color=white!15!black] table[row sep=crcr] {%
0.507692307692308	0\\
5.49230769230769	0\\
};
\addlegendentry{CVX Rand \cite{sidiropoulos}}

\addplot[ybar, bar width=0.123, fill=pink, draw=black, area legend] table[row sep=crcr] {%
1	1.56519553903691\\
2	1.56225357987456\\
3	1.58661014545333\\
4	1.563999789654\\
5	2.11674056571617\\
};
\addplot[forget plot, color=white!15!black] table[row sep=crcr] {%
0.507692307692308	0\\
5.49230769230769	0\\
};
\addlegendentry{ADMM \cite{bjornson}}

\addplot[ybar, bar width=0.123, fill=yellow, draw=black, area legend] table[row sep=crcr] {%
1	1.5687002283343924\\
2	1.5687098119844062\\
3	1.5688918477645351\\
4	1.555064255046005\\
5	1.94358686347932\\
};
\addplot[forget plot, color=white!15!black] table[row sep=crcr] {%
0.507692307692308	0\\
5.49230769230769	0\\
};
\addlegendentry{SNR Inc. \cite{hunger2007design}}
\end{axis}
\end{tikzpicture}%
}
        \caption{SNR of Users for $M=10, K=5$ at 10 dB}
        \label{fig2}
    \end{subfigure}

    \vspace{0.4cm}

    \begin{subfigure}{0.48\textwidth}
        \centering
        \scalebox{0.8}{\definecolor{mycolor1}{rgb}{1.00000,1.00000,0.00000}%
\definecolor{mycolor2}{rgb}{1.00000,0.00000,1.00000}%
\definecolor{mycolor3}{rgb}{0.00000,1.00000,1.00000}%
\begin{tikzpicture}

\begin{axis}[%
width=0.9\figW,
height=\figH,
at={(0\figW,0\figH)},
scale only axis,
xmin=0,
xmax=30,
xlabel style={font=\color{white!15!black}},
xlabel={$\bf{Transmit\:Power\:{\it{P_t}}\:in\:dB}$},
ymin=0,
ymax=70,
ylabel style={font=\color{white!15!black}},
ylabel={$\bf{Min-SNR}$},
axis background/.style={fill=white},
xmajorgrids,
ymajorgrids,
legend style={at={(0,0.8)}, anchor=south west, legend cell align=left, align=left, draw=white!15!black, row sep=-0.05cm, font=\scriptsize},
legend columns=3
]

\addplot [color=blue,  dashdotted,line width=1.0pt, mark=o, mark options={solid, blue}]
  table[row sep=crcr]{%
0	0.0520626018501649\\
10	0.519726204503723\\
20	5.19820588358208\\
30	51.9793587175767\\
};
\addlegendentry{CVX Rand \cite{sidiropoulos}}

\addplot [color=red,  line width=1.0pt]
  table[row sep=crcr]{%
0	0.0527783358945249\\
10	0.543210951649917\\
20	5.44856988808823\\
30	54.8797726114019\\
};
\addlegendentry{Algo RB}

\addplot [color=green, dashdotted, line width=1.0pt, mark=o, mark options={solid, green}]
  table[row sep=crcr]{%
0	0.0686912709669132\\
10	0.6869127363011\\
20	6.8691273891578\\
30	68.6912737651748\\
};
\addlegendentry{CVX Relaxed \cite{sidiropoulos}}

\addplot [color=yellow,  line width=1.0pt, mark=o, mark options={solid, yellow}]
  table[row sep=crcr]{%
0	0.0457020179728426\\
10	0.457020179728426\\
20	4.57020179728426\\
30	49.7020179728426\\
};
\addlegendentry{SNR Inc. \cite{hunger2007design}}

\addplot [color=pink, dashdotted, line width=1.0pt, mark=star, mark options={solid, pink}]
  table[row sep=crcr]{%
0	0.0429971881893984\\
10	0.456650568206385\\
20	4.63138660072947\\
30	49.7792677675973\\
};
\addlegendentry{ADMM \cite{bjornson}}
\end{axis}
\end{tikzpicture}%
}
        \caption{Min-SNR vs $P_t$ for $M=8, K=10$}
        \label{fig3}
    \end{subfigure}
    \hfill
    \begin{subfigure}{0.48\textwidth}
        \centering
        \scalebox{0.6}{
%
%
\definecolor{mycolor1}{rgb}{0.00000,0.44700,0.74100}%
\definecolor{mycolor2}{rgb}{0.85000,0.32500,0.09800}%
\definecolor{mycolor3}{rgb}{0.92900,0.69400,0.12500}%
\definecolor{mycolor4}{rgb}{0.49400,0.18400,0.55600}%
\definecolor{mycolor5}{rgb}{0.46600,0.67400,0.18800}%
\begin{tikzpicture}

\begin{axis}[%
 width=1.3\columnwidth,   
height=0.5\columnwidth,  
at={(2.6in,1.094in)},
scale only axis,
bar shift auto,
xmin=0.507692307692308,
xmax=10.4923076923077,
xtick={ 1,  2,  3,  4,  5,  6,  7,  8,  9, 10},
xlabel style={font=\color{white!15!black}},
xlabel={\textbf{User Index}},
ymin=0,
ymax=1.2,
ylabel style={font=\color{white!15!black}},
ylabel={$\bf{SNR}$},
axis background/.style={fill=white},
axis background/.style={fill=white},
xmajorgrids=true,          
ymajorgrids=true,
grid style={line width=0.8pt, draw=gray!60}, 
legend style={at={(0,0.970)}, anchor=south west, legend cell align=left, align=left, draw=white!15!black, row sep=-0.05cm, font=\scriptsize},
legend columns=5
]

\addplot[ybar, bar width=0.123, fill=red, draw=black, area legend] table[row sep=crcr]  {%
1	0.687326511093117\\
2	0.687326511090558\\
3	0.687326511090427\\
4	0.687326511095311\\
5	0.687326511088856\\
6	0.687326511089419\\
7	0.687326511089065\\
8	0.687326511094539\\
9	0.687326511094628\\
10	0.687326511101967\\
};
\addplot[forget plot, color=white!15!black] table[row sep=crcr] {%
0.507692307692308	0\\
10.4923076923077	0\\
};
\addlegendentry{Algo RB}

\addplot[ybar, bar width=0.123, fill=green, draw=black, area legend] table[row sep=crcr]  {%
1	0.693963677777486\\
2	0.693963709754177\\
3	0.693963660031019\\
4	0.693963660425567\\
5	0.693963676285373\\
6	0.693963659867869\\
7	0.693963662190605\\
8	0.693963659800214\\
9	0.693963661234317\\
10	0.69396365949085\\
};
\addplot[forget plot, color=white!15!black] table[row sep=crcr] {%
0.507692307692308	0\\
10.4923076923077	0\\
};
\addlegendentry{CVX Relaxed \cite{sidiropoulos}}

\addplot[ybar, bar width=0.123, fill=blue, draw=black, area legend] table[row sep=crcr]{%
1	0.755936916456863\\
2	0.634398210625692\\
3	0.591510170752328\\
4	0.596594813209476\\
5	0.821356184514743\\
6	0.782169943404099\\
7	0.715388359666981\\
8	0.7774291293654\\
9	0.590864844223472\\
10	0.700164319167997\\
};
\addplot[forget plot, color=white!15!black] table[row sep=crcr] {%
0.507692307692308	0\\
10.4923076923077	0\\
};
\addlegendentry{CVX Rand \cite{sidiropoulos}}

\addplot[ybar, bar width=0.123, fill=pink, draw=black, area legend] table[row sep=crcr]  {%
1	0.780062050474476\\
2	1.0660185660136\\
3	0.500119109995786\\
4	0.66732844305906\\
5	0.953002455288105\\
6	0.559610152991373\\
7	0.816052571834611\\
8	0.487543252167117\\
9	0.763814994370542\\
10	0.465492036079098\\
};
\addplot[forget plot, color=white!15!black] table[row sep=crcr] {%
0.507692307692308	0\\
10.4923076923077	0\\
};
\addlegendentry{ADMM \cite{bjornson}}

\addplot[ybar, bar width=0.123, fill=yellow, draw=black, area legend] table[row sep=crcr] {%
1	0.6247985252526696\\
2	0.5141266455188961\\
3	0.4431867668152448\\
4	0.845412982835787\\
5	0.5225917601345537\\
6	0.4275240380086877\\
7	0.6208339343863813\\
8	0.4214095438974732\\
9	0.493678178823867\\
10	0.5192182733585474\\
};
\addplot[forget plot, color=white!15!black] table[row sep=crcr] {%
0.507692307692308	0\\
10.4923076923077	0\\
};
\addlegendentry{SNR Inc. \cite{hunger2007design}}
\end{axis}

\end{tikzpicture}%
}
        \caption{SNR of Users for $M=8, K=10$ at 10 dB}
        \label{fig4}
    \end{subfigure}

    \caption{SNR comparison in underloaded and overloaded scenarios.}
    \label{fig:all}
\end{figure*}
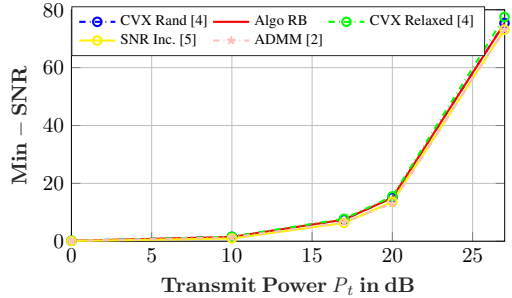
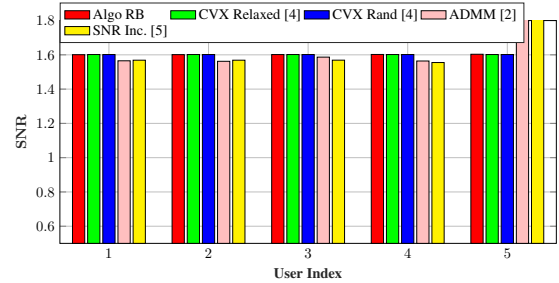
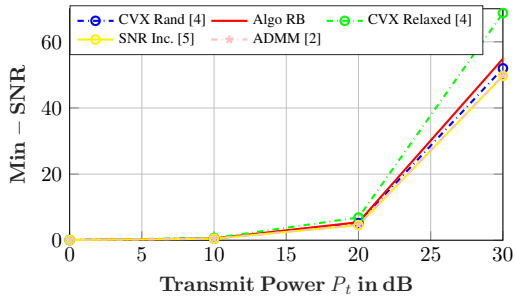
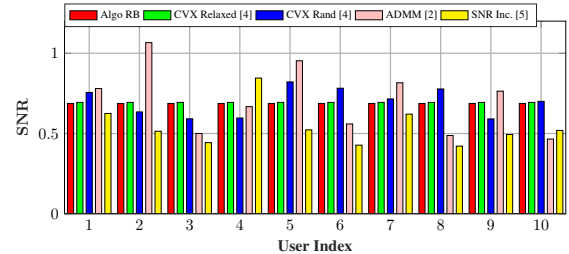
\section{Convergence and Complexity}
The original problem \eqref{P1} is non-convex and NP-hard. We resort to the FP approach of \cite{fractional1, fractional2} to reformulate \eqref{P1} to \eqref{P3} using auxiliary variables $\beta_k$'s, thereby making \eqref{P3} block-wise convex w.r.t. $\bfw$ and $\beta_k$'s separately. Since each subproblem is convex and its global optimal solution can be computed in closed-form, the convergence of the algorithm is guaranteed \cite{bcd}. The complexity of the proposed algorithm is dominated by computing the inverse of the $K \times K$ matrix $\bfD$ in \eqref{sol1}, which is significantly cheaper than the state-of-the-art algorithms such as \cite{sidiropoulos, bjornson2} that employ CVX solvers with the SDR approach.
\section{Results and Discussion}
 In this section, numerical results are provided to validate the effectiveness of the proposed algorithm. The channels are generated according to the 3GPP specifications \cite{etsi5138}, as discussed in \cite{syed2025bilinear}. The proposed rate balancing algorithm {\it{Algo RB}} is compared with the SDR algorithm of \cite{sidiropoulos}, denoted by  {\it{CVX Rand} } and its rank-relaxed solution {\it{CVX Relaxed}}. We also compare the performace with other non-CVX based algorithms {\it{ADMM}} \cite{bjornson} and {\it{SNR Inc.}} \cite{hunger2007design}. The min SNR of the users is averaged over 500 channel realisations. We assume the channels of $K$ users to be independent of each other. We first plot the minimum SNR of the users versus transmit power $P_t$ in dB for $M = 10$ and $K = 5$ in Fig.~\ref{fig1}. All the algorithms perform close to each other in this underloaded scenario. Next, we compare the SNR of different users for $P_t = 10$~dB in Fig.~\ref{fig2}. The figure shows that the relaxed CVX solution, which is the upper bound for the MMF problem leads to balanced rates of all users. In this setup, {\it{CVX Rand}} (computed with 10,000 randomisations) also leads to balanced rates of all users. The proposed algorithm performs similar to {\it{CVX Relaxed}} and {\it{CVX Rand}}, and slightly better than {\it{ADMM}} and {\it{SNR Inc.}} The solutions delivered by {\it{ADMM}} and {\it{SNR Inc.}} achieve near optimal performance but they do not result in balanced user rates, as seen in Fig.~\ref{fig2}. This confirms that balanced rates yield the best performance when $K \leq M$, as proved earlier.
We next consider the overloaded scenario with $M = 8$ and $K = 10$ in Fig.~\ref{fig3}. Even though $K > M$ here, the considered system scenario results in a full rank matrix $\bfD$, and so, there is no need to search for the active set of users, and \eqref{sol1} is directly applicable. As  seen from Fig.~\ref{fig3}, the proposed algorithm performs the best among all algorithms and it even outperforms the high complexity {\it{CVX Rand}}. For this overloaded scenario, the upper bound given by the relaxed CVX solution is not tight at the higher transmit power regimes. We also analyse the SNR of different users for this setup at $P_t = 10$~dB in Fig.~\ref{fig4}. The figure shows that the upper bound of the solution, i.e., {\it{CVX Relaxed}} results in balanced SNRs of all the users. However, {\it{CVX Rand}} obtained after randomisation, results in unbalanced SNRs, and its minimum SNR is lower than the balanced SNR obtained by the proposed algorithm. This plot once again confirms our theory that rate balancing leads to the optimal MMF multicasting solution as long as the matrix $\bfD$ of the proposed algorithm is invertible.    
\section{Conclusion}
In this work, we have proved that rate balancing is the optimal solution for MMF multicasting when $K \leq M$ and the channels are linearly independent. This setup is common in practice due to the ongoing demand for massive MIMO systems. The proposed algorithm, built on this insight, is computationally efficient compared to the CVX-based solutions. Further analysis for $K > M$ will be presented in our next work.
\FloatBarrier
\bibliographystyle{IEEEbib}
\bibliography{strings,refs}

\end{document}